\documentclass[pra,twocolumn,reprint]{revtex4-1}

\usepackage{amsmath,amssymb,amsthm,bm,braket,ascmac,bbm,color}
\usepackage{graphicx}

\newcommand{\invS}{\hat{\mathcal{I}}_s}
\newcommand{\invB}{\hat{\mathcal{I}}_b}

\newcommand{\hext}{\hat{H}_\text{ext}}
\newcommand{\htot}{\hat{H}_\text{tot}}
\newcommand{\hzero}{\hat{H}_0}

\newcommand{\hdimer}{\hat{H}_\text{dimer}}
\newcommand{\hdimermat}{H_\text{dimer}}
\newcommand{\gs}{\ket{\text{gs}}}

\newcommand{\hS}{\hat{S}}

\newcommand{\hJ}{\hat{J}_\text{spin}}
\newcommand{\hlocj}{\hat{j}^\text{spin}}

\newcommand{\amp}{\alpha}
\newcommand{\ampp}{\beta}
\newcommand{\tfwhm}{T_\text{FWHM}}
\newcommand{\tini}{t_\text{ini}}

\newcommand{\cre}{\hat{c}^\dag}
\newcommand{\ann}{\hat{c}}
\newcommand{\spinor}{\psi_{k}}

\newcommand{\hzeromat}{H_0}
\newcommand{\hextmat}{H_\text{ext}}

\newcommand{\Jspinmat}{J_\text{spin}}
\newcommand{\Jpulse}{J'_{\text{pulse}}}

\newcommand{\epI}{\epsilon(k)}
\newcommand{\gapI}{\Delta}

\newcommand{\stagXY}{\eta_\text{stag}}
\newcommand{\stagZ}{H_\text{stag}}

\newcommand{\bj}{\mathcal{J}}
\newcommand{\omu}{\bar{\Omega}}

\newcommand{\dd}{\mathrm{d}}
\newcommand{\ee}{\mathrm{e}}
\newcommand{\ii}{\mathrm{i}}

\newcommand{\hl}[1]{\textcolor{black}{#1}}

\begin{document}
\title{Generation of dc, ac, and second-harmonic spin currents\\ by electromagnetic fields in an inversion-asymmetric antiferromagnet}
\author{Tatsuhiko N. Ikeda}
\affiliation{Institute for Solid State Physics, University of Tokyo, Kashiwa, Chiba 277-8581, Japan}
\date{\today}
\begin{abstract}
Manipulating spin currents in magnetic insulators is a key technology in spintronics.
We theoretically study a simple inversion-asymmetric model of quantum antiferromagnets,
where both the exchange interaction and the magnetic field are staggered.
We calculate spin currents generated by external electric and magnetic fields by using a quantum master equation.
We show that an ac electric field with amplitude $E_0$ leads, through exchange-interaction modulation, to the dc and second-harmonic spin currents proportional to $E_0^2$.
We also show that dc and ac staggered magnetic fields $B_0$ generate the dc and ac spin currents proportional to $B_0$, respectively.
We elucidate the mechanism by an exactly solvable model,
and thereby propose the ways of spin current manipulation by electromagnetic fields.
\end{abstract}
\maketitle
\section{Introduction}

Spintronics has attracted growing attention in fundamental and applied physics
for decades~\cite{Zutic2004,Kirilyuk2010,Maekawa2017},
where the researchers have explored how to 
manipulate the spin degree of freedom in materials and devices~\cite{Sinova2012}.
For example, the spin Hall effect deriving from the spin-orbit coupling
enables the conversion between the spin current and the electric current~\cite{Dyakonov1971,Hirsch1999,Kato2004},
and the spin Seebeck effect~\cite{Uchida2008} extends to the research field of spin caloritronics~\cite{Bauer2012}.
One important class of materials in spintronics is the magnetic insulator,
where the charge degree of freedom is frozen and magnetic excitations play the principal role~\cite{Chumak2015}.
Being free from Ohmic losses,
the spin currents in these materials are expected to be useful for future computing devices~\cite{Chumak2014}.
Thus, it has been of crucial importance to develop the ways to control these spin currents freely~\cite{Krawczyk2014}.

Antiferromagnets have emerged as a new class of materials
whose unique features 
have turned out to be suited for spintronic applications~\cite{Baltz2018}.
\hl{For example, the time scale of magnetic excitations
of antiferromagnets is typically shorter than that of ferromagnets,
the antiferromagnets are promising candidates for high-speed spintronic devices}~\cite{Kimel2004}.
Among several approaches including thermal effects~\cite{Seki2015,Lin2016,Naka2019},
the optical control of antiferromagnets,
\hl{which enables the fastest manipulation,}
has attracted considerable attention~\cite{Kimel2005,Satoh2007,Zhou2012,Nishitani2013,Mukai2014}.
Recently, Ishizuka and Sato~\cite{Ishizuka2019a,Ishizuka2019b} 
have theoretically shown that inversion-asymmetric antiferromagnets are useful for spin-current generation
by electromagnetic waves.
They have proposed the spin-current rectification in ac electric and magnetic fields,
where the magnitude of the generated dc spin current is proportional to the second power
of the input-field amplitude.
The dc spin-current generation as rectification has been also numerically confirmed
and the second-harmonic spin current is studied in Ref.~\cite{Ikeda2019}.

In this paper, we propose two other ways to produce spin currents
by electromagnetic fields in inversion-asymmetric antiferromagnets.
We consider a one-dimensional model for them, where both the exchange interaction
and the magnetic field are staggered,
and study the spin current induced by an electric or magnetic field of pulse shape
by numerically integrating a quantum master equation.
On one hand, we show that an ac electric field of amplitude $E_0$ leads to exchange-interaction modulation~\cite{Mentink2015}
and gives rise to the dc and second-harmonic spin currents
whose magnitude are proportional to $E_0^2$.
\hl{Note that this type of coupling between the spin system
and the electric field is generic, and thus not restricted
to multiferroic systems}~\cite{Ishizuka2019a,Ishizuka2019b,Ikeda2019}.
On the other hand, we show that dc and ac staggered magnetic fields of amplitude $B_0$
generate the dc and ac spin currents, respectively, whose magnitude are both proportional to $B_0$.
The underlying mechanism of these spin current generations are elucidated in a unified manner
as the competition between the staggered exchange interaction and magnetic field.
\hl{This mechanism is distinct from the spin-current rectification proposed in Refs.}~\cite{Ishizuka2019a,Ishizuka2019b}.

\section{Formulation of the Problem}\label{sec:formulation}

\subsection{Time-independent Hamiltonian}
In this work, we consider the following Hamiltonian for a spin chain~\cite{Ishizuka2019a}
\begin{align}
	\hzero &=  \sum_{j=1}^{2L} \left\{J\left[ 1+ (-1)^j   \stagXY \right] (\hS_j^x \hS_{j+1}^x + \hS_j^y \hS_{j+1}^y)\right.\notag\\
		&\qquad\qquad\qquad \left. +(-1)^j\stagZ\hS_j^z\right\}.\label{eq:H0}
\end{align}
Here $\hS_j^\alpha$ ($\alpha=x,y$ and $z$) denote the spin operators at site $j$
for the spin-1/2 representation,
$J$ $(>0)$ is the exchange interaction,
and $\stagXY$ ($\stagZ$) is the staggered exchange interaction (magnetic field).
This model is useful to study inversion-asymmetric antiferromagnets
(see, e.g., Ref.~\cite{Ikeda2019} and references therein for the candidate materials).
We impose the periodic boundary conditions $\hS_{2L+j}^\alpha=\hS_j^\alpha$.

\begin{figure}
\centering
\includegraphics[width=6cm]{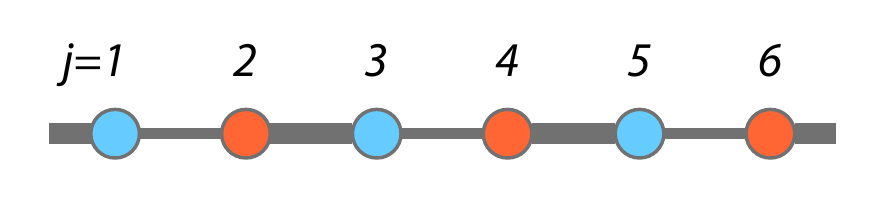}
\caption{
Schematic illustration of our model~\eqref{eq:H0}.
The thick and thin bonds represent the exchange couplings $J(1+\stagXY)$ and $J(1-\stagXY)$, respectively,
and the red and blue sites do the local magnetic fields $\stagZ$ and $-\stagZ$, respectively.
}
\label{fig:model}
\end{figure} 

There are two kinds of inversion transformation regarding this model:
the site-center inversion $\invS$ and the bond-center inversion $\invB$.
These inversions are characterized, for instance, by
$\invS\hS_j^\alpha\invS^\dag = \hS_{-j}^\alpha$
and $\invB\hS_j^\alpha\invB^\dag = \hS_{-j+1}^\alpha$.
It follows from Eq.~\eqref{eq:H0} that
\begin{align}
\invS\hzero(\stagXY,\stagZ)\invS^\dag &=\hzero(-\stagXY,\stagZ),\notag\\
\invB\hzero(\stagXY,\stagZ)\invB^\dag &=\hzero(\stagXY,-\stagZ).	\label{eq:H0symm}
\end{align}
Thus, our Hamiltonian is symmetric under the site-center inversion for $\stagXY=0$
and under the bond-center inversion for $\stagZ=0$.
When neither $\stagXY$ nor $\stagZ$ vanishes, our Hamiltonian is inversion-asymmetric.
As we will see below, the spin current arises only for the inversion-asymmetric situation
in our setup.

\subsection{Coupling to ac electric field: difference- and sum-frequency mechanisms}
We suppose that our spin model is a low-energy effective model
of strongly correlated electrons.
Specifically, we regard the exchange interaction $J$ as
a superexchange of the one-dimensional Hubbard model at half filling
with transfer integral $t_0$ and on-site Coulomb interaction $U$.
Then we obtain $J=4t_0^2/U$~\cite{Anderson1950}.

Now we consider the effect of an ac \textit{electric} field along the spin chain.
Although the spin chain apparently does not couple to the electric field,
it does through virtual hopping processes of the underlying charge degrees of freedom
in the Hubbard model~\cite{Mentink2015}. As shown in Refs.~\cite{Kitamura2017,Chinzei},
the ac electric field makes the exchange interaction $J$ be time-dependent as
\begin{align}\label{eq:Jt_raw}
	J(t) = \sum_{m,n}(-1)^m\frac{4t_0^2 \bj_{n+m}(F)\bj_{n-m}(F)}{U-(n+m)\Omega} \cos(2m\Omega t),
\end{align}
where $\bj_n(x)$ is the Bessel function of the first kind and
$\Omega$ the angular frequency of the ac electric field.
The dimensionless parameter $F=eaE_0/\Omega$ ($\hbar=1$ throughout this paper) represents the coupling strength
between the electron and the ac electric field, where $e$ $(>0)$ is the elementary charge,
$a$ the lattice constant, and $E_0$ is the field amplitude.

Let us assume that $F\ll1$ and simplify Eq.~\eqref{eq:Jt_raw}.
Under this condition, we have $\bj_{n+m}(F)\bj_{n-m}(F)=O(F^{|n+m|+|n-m|})$,
which implies that the coefficient of $\cos(2m\Omega t)$ is $O(F^{m})$
and the higher frequency component rapidly decreases.
Thus we ignore the terms with $|m|\ge2$ in Eq.~\eqref{eq:Jt_raw},
obtaining
\begin{align}\label{eq:Jt_app}
	J(t) &\simeq J + J'(t),\\
	J'(t) &= J\frac{F^2}{2}\frac{\omu^2}{1-\omu^2} - J\frac{F^2}{2}\frac{\omu^2(1+2\omu^2)}{(1-\omu^2)(1-4\omu^2)}\cos(2\Omega t),\label{eq:Jp1}
\end{align}
where $\omu\equiv \Omega/U$ and we have ignored higher-order correction terms in $F$.
We assume $\omu<1$ throughout this paper, and further simplify Eq.~\eqref{eq:Jp1} as
\begin{align}
	J'(t) \simeq J\frac{F^2}{2} \omu^2\left[ 1- \cos(2\Omega t)\right] = J \amp \sin^2(\Omega t)\label{eq:Jp2}
\end{align}
with
\begin{align}
	\amp \equiv F^2\omu^2 = \left( \frac{eaE_0}{U}\right)^2.\label{eq:ampdef}
\end{align}
Here we have ignored higher-order correction of $O(\omu^4)$.

We emphasize that the frequencies involved in the exchange interaction~\eqref{eq:Jp2} are $0\Omega$ and $2\Omega$
rather than $\Omega$ of the applied ac field.
These are kinds of difference-frequency $(\Omega-\Omega)$
and sum-frequency $(\Omega+\Omega)$ generation.
The exchange interaction modulation in the spin model
derives from the second-order virtual processes of the underlying charge degrees of freedom.
In fact, the amplitude $\alpha$ of the exchange interaction modulation is proportional to $E_0^2$ as in Eq.~\eqref{eq:ampdef}.
Thus, as we will show below, the linear response as a spin model to the exchange interaction modulation $\alpha$
gives rise to the dc ($0\Omega$) and second-harmonic ($2\Omega$) outputs.

\subsection{Total Hamiltonian and spin current}
We complete the formulation of the problem that we address in this paper.
Combining the above arguments,
we arrive at the following spin-system Hamiltonian
\begin{align}
	\htot(t) &= \hzero + \hext(t),\label{eq:Htot}\\
	\hext(t) &= J'(t) \sum_{j=1}^{2L} \left[ 1+ (-1)^j   \stagXY \right] (\hS_j^x \hS_{j+1}^x + \hS_j^y \hS_{j+1}^y),\label{eq:Hext}
\end{align}
where $\hzero$ is defined in Eq.~\eqref{eq:H0}.

Note that the total Hamiltonian $\htot(t)$ has the global U(1) symmetry associated with the rotation around the $S^z$ axis.
Thus the total magnetization $\hat{M}^z\equiv \sum_j \hat{S}^z_j/2$ is conserved,
and the continuity equations for local $\hat{S}^z$'s hold true: $d \hat{S}^z_i/dt + \hlocj_i - \hlocj_{i-1}=0$.
Here $\hlocj_j\equiv[J+J'(t)] [1+(-1)^j \stagXY ] (\hS^x_j \hS^y_{j+1}-\hS^y_j \hS^x_{j+1})$
represents the local spin current flowing between the sites $j$ and $j+1$.
The observable of interest is the total spin current 
\begin{align}
	\hJ &= \sum_j \hlocj_j \notag\\
	&=  [J+J'(t)]\sum_j [1+(-1)^j \stagXY ] (\hS^x_j \hS^y_{j+1}-\hS^y_j \hS^x_{j+1})\label{eq:sc}.
\end{align}
Since we focus on the case in which $J'(t)$ is small, we may safely neglect $J'(t)$ from Eq.~\eqref{eq:sc}.
In the following, we consider the ground state of $\hzero$
and analyze the spin current $\hJ$ generated by the time-dependent perturbation $\hext(t)$.

\hl{Note that the spin current is parallel to the ac electric field, which has been assumed to be along the chain.
If we applied the ac electric field perpendicular to the chain, the exchange interaction modulation would not happen and there would be no spin current generation in our one-dimensional model.
This is not true in general two-dimensional systems.
In fact, Naka et al. have recently shown that a dc electric field or thermal gradient leads to a spin current perpendicular to it in two-dimensional organic antiferromagnets}~\cite{Naka2019}. 
 
For later use,
we remark that $\hJ$ is odd under both inversions:
\begin{equation}
	\invS \hJ \invS^\dag =\invB \hS \invB^\dag  =-\hJ.\label{eq:scsymm}
\end{equation}
Therefore, when $\stagXY=0$ or $\stagZ=0$ and either inversion symmetry is present,
no dynamics occurs in the spin current.
For the spin current generated, both $\stagXY$ and $\stagZ$ must be nonzero.

\section{Results}


\subsection{Dc and second-harmonic spin currents}\label{sec:master}
\begin{figure*}
\centering
\includegraphics[width=15cm]{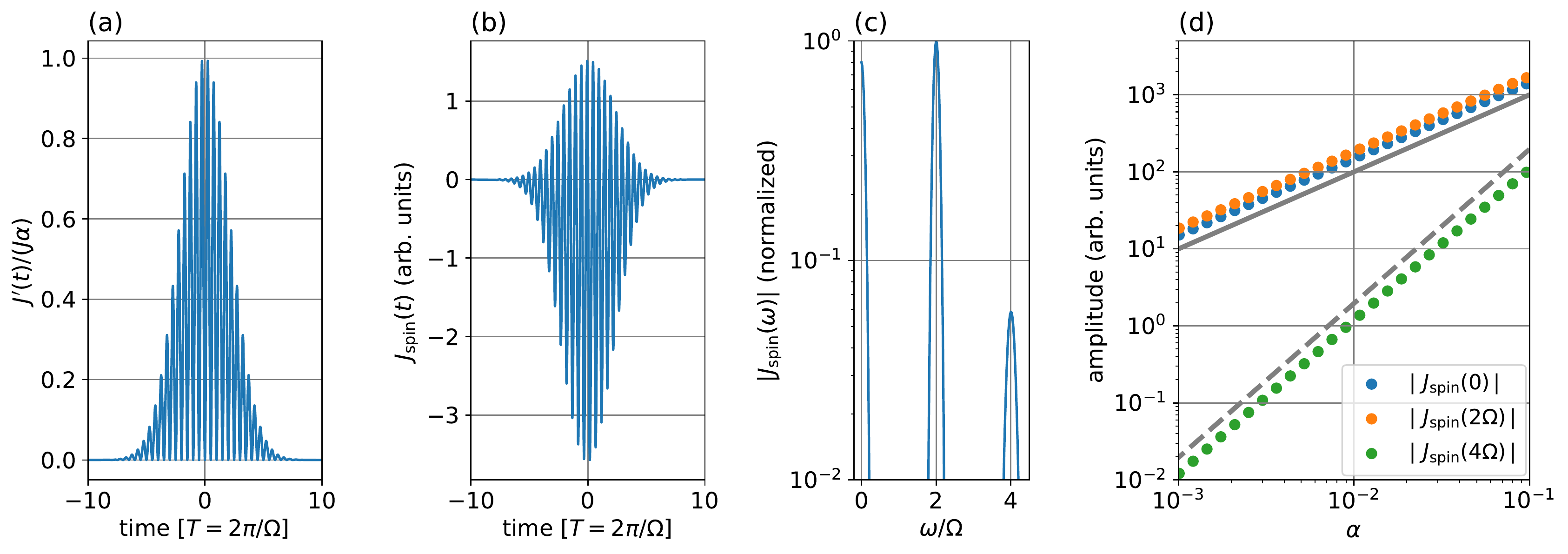}
\caption{
(\textbf{a}) Form of oscillating exchange interaction $\Jpulse$~\eqref{eq:Jpulse}.
(\textbf{b}) Time profile of spin current $\Jspinmat(t)$ for $(\stagXY,\stagZ/J)=(0.1,0.03)$
and $\Omega=5\times10^{-2}J$ and $\alpha=0.1$.
(\textbf{c}) Amplitude of the corresponding Fourier transform.
(\textbf{d}) Amplitudes of the dc and harmonic spin currents,
$|\Jspinmat(n\Omega)|$ ($n=0,2$, and 4), plotted against
the exchange-interaction-modulation amplitude $\alpha$.
The solid (dashed) line shows the slope 1 (2) for the guide to the eye.
}
\label{fig:master}
\end{figure*}

We now numerically investigate the spin current dynamics
under a multi-cycle pulse field of experimental interest.
We replace $J'(t)$ in Eq.~\eqref{eq:Jp2} by
\begin{equation}
	J'(t) \to \Jpulse(t) \equiv J \amp f(t) \sin^2(\Omega t),\label{eq:Jpulse}
\end{equation}
where the Gaussian envelope function
$f(t) = \exp[-4\ln 2\cdot (t/\tfwhm)^2]$ with full width at half maximum $\tfwhm$.
To be specific, we set $\tfwhm = 10\pi/\Omega$,
for which $\Jpulse(t)$ is illustrated in Fig.~\ref{fig:master}(a).
We have confirmed that the results are not sensitive to the pulse width.

The actual numerical calculations are as follows.
We take an initial time $\tini$ $(<0)$ so that $\Jpulse(\tini)$
is negligibly small,
and suppose that the spin system is in the ground state
with zero total magnetization $\hat{M}^z=\sum_j \hat{S}^z_j/2=0$.
Then we solve the dynamics represented by a quantum master equation
(see Sec.~\ref{sec:mate} for detail),
which describes the time-dependent Schr\"{o}dinger equation
in the presence of relaxation.
We set the relaxation rate as $\gamma=0.1J$.
\hl{Our master equation ensures that, without the external field, the system relaxes to
the ground state i.e. the zero-temperature state. Thus thermal fluctuations}~\cite{Seki2015} \hl{are neglected
in our model.}

Figure~\ref{fig:master}(b) shows a typical time profile of the spin current.
The Hamiltonian parameters are $(\stagXY,\stagZ/J)=(0.1,0.03)$
and the field parameters are $\Omega=5\times10^{-2}J$ and $\alpha=0.1$.
The panel (c) shows the corresponding Fourier spectrum $|\Jspinmat(\omega)|$,
which consists of the dc ($\omega=0$), second-harmonic $(\omega=2\Omega)$,
and forth-harmonic $(\omega=4\Omega)$ components.
As we emphasized in Sec.~\ref{sec:formulation},
there appear even-order harmonics because the exchange-interaction modulation~\eqref{eq:Jpulse} consists of the dc and second-harmonic components
and no longer involves the fundamental frequency $\Omega$ of the input laser.

The laser-intensity dependence of each harmonic spin current
is shown in Fig.~\ref{fig:master}(d).
In the log-log scale,
the dc and second-harmonic data follow a line with slope 1
whereas the fourth-harmonic ones with slope 2.
Therefore, the dc and second-harmonic components are proportional to $\alpha$.
Meanwhile, the fourth-harmonic component is proportional to $\alpha^2$
and, thus, arises from the second-order process in terms of $\alpha$.
In terms of the ac-electric-field amplitude $E_0$,
the dc and second-harmonic spin currents are $O(E_0^2)$
whereas the fourth-harmonic is $O(E_0^4)$.
Note that the fourth-harmonic spin current may become different
if we incorporate the terms with $m=\pm2$ in Eq.~\eqref{eq:Jt_raw},
which are $O(E_0^4)$ and neglected in our calculation.
However, the dc and second-harmonic spin currents are not affected much
by these higher-order terms.

\subsection{Direction of dc spin current}
In Sec.~\ref{sec:master}, we showed a typical behavior of the spin current
by fixing $(\stagXY,\stagZ/J)=(0.1,0.03)$.
Here we focus on the dc component and
study how its direction depends on the Hamiltonian parameters $\stagXY$ and $\stagZ$.

\begin{figure}
\centering
\includegraphics[width=6cm]{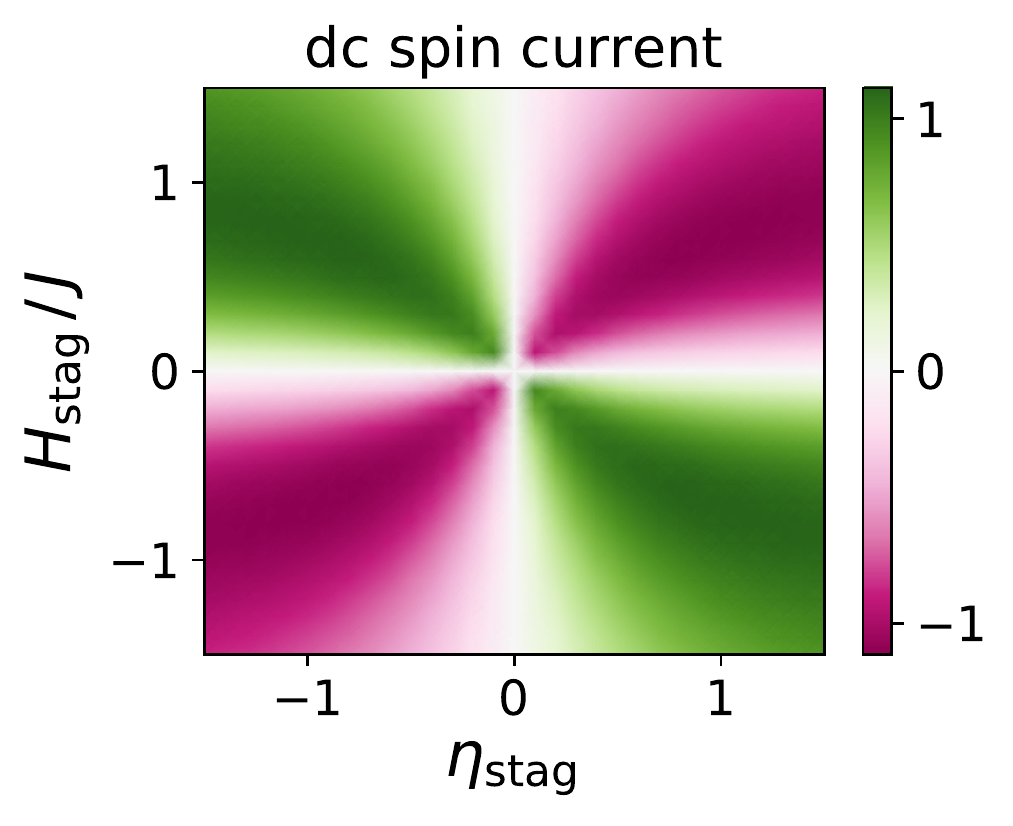}
\caption{
Rescaled dc spin current calculated in the pulse dynamics
over the $(\stagXY,\stagZ)$-plane.
The other parameters are set as
$\tfwhm = 10\pi/\Omega$, $\Omega=5\times10^{-2}J$ and $\alpha=0.1$.
}
\label{fig:color}
\end{figure} 

Figure~\ref{fig:color} shows the dc spin current
in the $(\stagXY,\stagZ)$-plane obtained
by the similar calculation in Sec.~\ref{sec:master}.
The color bar is renormalized by a positive scale factor for display.
Since we work in the linear response regime, the color map does not change
with $\alpha$ varied as long as $\alpha$ is sufficiently small.
In the first quadrant $\stagXY>0$ and $\stagZ>0$, the dc spin current is negative.
Note that this is consistent with Fig.~\ref{fig:master}(b),
where the dc component, or the time average of $\Jspinmat(t)$, is negative.

The dc component vanishes on the lines $\stagXY=0$ and $\stagZ=0$
and its magnitude increases as $(\stagXY,\stagZ)$ goes away from these lines.
On these lines, either the bond-center or the site-center inversion symmetry
arises,
and not only the dc component but also the total spin current $\Jspinmat(t)$
vanishes as shown in Sec.~\ref{sec:formulation}.

This inversion-symmetry argument explains why
Fig.~\ref{fig:color} is antisymmetric under reflections
across each of the $\stagXY$ and $\stagZ$ axes.
As shown in Eq.~\eqref{eq:H0symm},
the sign change of $\stagZ$ ($\stagXY$) is equivalent to
applying the site-center (bond-center) inversion.
On the other hand, each of the site-center and the bond-center inversion
changes the sign of the spin current as in Eq.~\eqref{eq:scsymm}.
From these two properties, it follows that
\begin{equation}
\Jspinmat^{-\stagXY,\stagZ}(t)=\Jspinmat^{\stagXY,-\stagZ}(t)=-\Jspinmat^{\stagXY,\stagZ}(t),
\end{equation}
and, hence, similar relations hold true for the dc components.

\subsection{Mechanism of spin current generation}\label{sec:mechanism}
\begin{figure*}
\centering
\includegraphics[width=15cm]{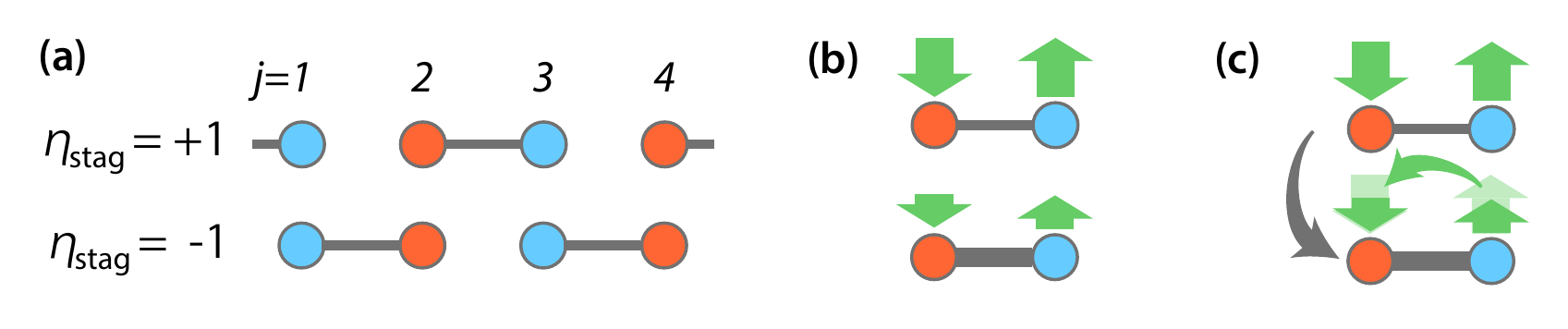}
\caption{
(\textbf{a}) Schematic illustration of our model~\eqref{eq:H0} for $\stagXY=1$ (upper)
and $\stagXY=-1$ (lower). The missing bonds show that the exchange couplings vanish on them.
(\textbf{b}) Isolated dimer~\eqref{eq:hdimer} for $\stagZ>0$,
where the red (blue) site shows the positive (negative) local magnetic field.
The green arrows represent the local magnetization on each site in the ground state~\eqref{eq:gs}.
The upper (lower) shows the dimer for the weaker (stronger) exchange interaction.
(\textbf{c}) Mechanism of spin current generation by exchange-interaction modulation.
As the exchange interaction increases, the local magnetization flows from right to left when $\stagZ>0$.
}
\label{fig:chains}
\end{figure*}

Here we look into the mechanism of the spin current generation
numerically obtained in the previous sections.
For this purpose, we focus on the special case of $\stagXY=1$,
for which the exchange interaction vanishes at every two bonds
and the spin chain is dimerized as illustrated in Fig.~\ref{fig:chains}(a).
Thus our spin chain problem reduces to a single dimer.

To investigate the ground state of a dimer,
we take $j=2$ and 3, for example.
Writing the spin operators as $\hS_L^\alpha=\hS_2^\alpha$ and $\hS_R^\alpha=\hS_3^\alpha$
($\alpha=x,y$, and $z$),
we have the following Hamiltonian:
\begin{equation}
	\hdimer = 2J (\hS_L^x\hS_R^x+\hS_L^y\hS_R^y)+\stagZ \hS_L^z - \stagZ \hS_R^z.\label{eq:hdimer}
\end{equation}
Among the 4 states $\ket{\uparrow\uparrow},\ket{\uparrow\downarrow},\ket	{\downarrow\uparrow}$, and $\ket{\downarrow\downarrow}$,
each of $\ket{\uparrow\uparrow}$ and $\ket{\downarrow\downarrow}$ is the eigenstate of $\hdimer$ with zero eigenvalue.
On the other hand, $\ket{\uparrow\downarrow}$ and $\ket{\downarrow\uparrow}$ couple to each other.
In the basis of these states, $\hdimer$ is represented by the $2\times2$ matrix:
\begin{equation}
	\hdimermat = \begin{pmatrix}
		\stagZ & J \\ J & -\stagZ
	\end{pmatrix},
\end{equation}
whose eigenvalues are $\pm\sqrt{J^2+\stagZ^2}$.
Thus the ground state $\gs$ is the eigenstate with negative eigenvalue
and given by
\begin{equation}
	\gs = -\sin\frac{\theta}{2}\ket{\uparrow\downarrow}+\cos\frac{\theta}{2} \ket{\downarrow\uparrow};\quad \cos\theta = \frac{\stagZ}{\sqrt{J^2+\stagZ^2}}.\label{eq:gs}
\end{equation}

The local magnetization distribution in $\gs$ follows from the exact solution:
\begin{equation}
	\braket{\text{gs}|\hS^z_L |\text{gs}} = -\frac{\stagZ}{2\sqrt{J^2+\stagZ^2}}
	= -\braket{\text{gs}|\hS^z_R |\text{gs}}.\label{eq:gsmag}
\end{equation}
For the special case of $\stagZ=0$, the local magnetization on both sites vanish
and there is no magnetization imbalance between the sites.
This is a manifestation of the bond-center inversion symmetry.
For $\stagZ>0$, the local magnetization on the left (right)
is negative (positive), and this situation is illustrated in Fig.~\ref{fig:chains}(b).
These signs become opposite for $\stagZ<0$.

As the exchange interaction $J$ increases,
the magnetization imbalance between the sites becomes smaller
as illustrated in Fig.~\ref{fig:chains}(b).
This tendency can be read from Eq.~\eqref{eq:gsmag}.
Also, we can make an intuitive interpretation as follows.
In the limit of $J\to0$, there is no spin exchange
and $\gs=\ket{\downarrow\uparrow}$ for $\stagZ>0$
so that the system maximizes each local magnetization and minimizes the energy.
As $J$ is turned on, the system decreases its energy further by using
the spin exchange, where the local magnetizations are decreased.
In fact, in the limit $J\to\infty$, the ground state becomes the spin singlet pair,
which has no local magnetization.
There exists a competing effect between $J$ and $\stagZ$:
$J$ prefers the spin singlet and less local magnetizations
and $\stagZ$ does larger magnetization imbalance.

From the above argument, we arrive at the understanding the spin current generation.
For $\stagZ>0$, the increase of $J$ corresponds to the transition from the upper to the lower pictures in Fig.~\ref{fig:chains}(b).
Here the magnetization flows from right to left in total,
or the dc spin current is negative as illustrated in the panel (c).
This explains why the first quadrant of Fig.~\ref{fig:color} gave negative values.
Note that the continuity equation for magnetization does not hold true exactly, unlike that for electric charge,
owing to the dissipation as shown in Sec.~\ref{sec:mate}.

It is now clear that the dc spin current is positive for $\stagXY<0$.
In this case, the signs of local magnetic fields
and, hence, local magnetic moments become opposite in Figs.~\ref{fig:chains}(b) and (c),
and the direction of the dc spin current is thus flipped.
Furthermore, it is also clear that the dc spin current changes its sign
if $\stagXY$ is changed from $+1$ to $-1$.
This change leads to the other parings of dimers as shown in the lower picture in Fig.~\ref{fig:chains}(a).
Here the local magnetic field on the left and right sites of the dimer
is flipped and, thus, the dc spin current changes its sign.

These interpretations are basically true for the general case $\stagXY\neq\pm1$.
Unless $\stagXY=0$, the exchange interaction is alternating.
Focusing on a bond with the stronger exchange,
we regard the two sites on the bond forming a dimer.
Unless $\stagZ=0$, the local magnetic fields on the two sites are different
and some magnetization imbalance exists in the dimer.
Then the increase of exchange interaction decreases the magnetization imbalance
and the dc spin current arises accordingly.

\subsection{dc spin current generation by external magnetic field}
\begin{figure*}
\centering
\includegraphics[width=15cm]{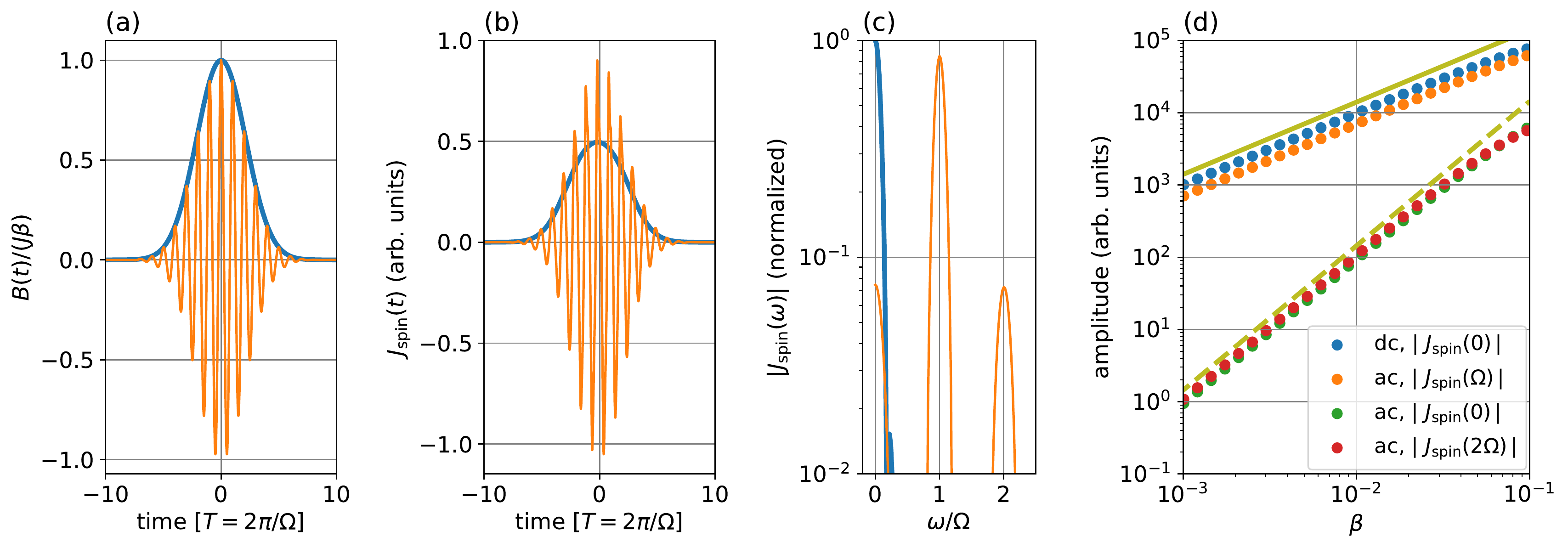}
\caption{
(\textbf{a}) Form of external magnetic field for the dc (blue) and ac (orange) cases.
(\textbf{b}) Time profile of spin current $\Jspinmat(t)$ for the dc (blue) and ac (orange) cases.
The parameters are $(\stagXY,\stagZ/J)=(0.1,0.03)$,
$\Omega=5\times10^{-2}J$, and $\beta=0.1$.
(\textbf{c}) Amplitude of the corresponding Fourier transform for the dc (blue) and ac (orange) cases.
The amplitude is normalized so that that $|\Jspinmat(\omega=0)|=1$ for the dc case.
(\textbf{d}) Amplitudes of the dc and harmonic spin currents plotted against
the magnetic-field amplitude $\beta$.
The blue show the dc spin current $|\Jspinmat(0)|$,
and the orange, green, and red show
$|\Jspinmat(n\Omega)|$ with $n=1,0$, and 2), respectively.
The solid (dashed) line shows the slope 1 (2) for the guide to the eye.
}
\label{fig:mag_master}
\end{figure*}

The spin-current-generation mechanism elucidated in the previous section is the
competing effect between the exchange couplings and the local magnetic fields.
This implies that the spin currents can also be generated by staggered magnetic fields.
To show this, we replace $\hext(t)$ discussed so far by the following term:
\begin{align}
	\hext'(t) &= B(t) \sum_{j=1}^{2L}(-1)^j\hS_j^z,\notag\\
	B(t) &\equiv J \ampp f(t)\cdot\begin{cases}
		1 & (\text{dc case}),\\
		\cos(\Omega t) & (\text{ac case}),
	\end{cases} 
	\label{eq:Hext2}
\end{align}
where $f(t)$ is the Gaussian envelope function defined below~\eqref{eq:Jpulse}.
For the ac case, we use $\Omega=5\times10^{-2}J$ and $\tfwhm=10\pi/\Omega=2\pi\times10^2 J^{-1}$ as in the previous sections.
For the dc case, we use the same pulse width 
$\tfwhm=2\pi\times10^2 J^{-1}$.
The forms of $B(t)$ for the dc and ac cases are shown in Fig.~\ref{fig:mag_master}(a).
The ac case has been studied in Refs.~\cite{Ishizuka2019a,Ikeda2019},
and we will compare the dc case results to it below.

In the experimental viewpoint,
the external magnetic field may be spatially uniform.
Since we consider the situation where internal staggered magnetic fields $\stagZ$ are present,
the external field induces the staggered component~\eqref{eq:Hext2} in general~\cite{Ishizuka2019a}.
Note that the uniform component of the external magnetic field causes
no physical effect since the total magnetization is a conserved quantity in our model.

Figures~\ref{fig:mag_master}(b) shows
the time profiles of the spin currents
for the dc and ac cases.
For the dc case, the generated spin current is positive
in contrast to the exchange-interaction modulation in Fig.~\ref{fig:master}(b).
This is in perfect agreement with the physical mechanism found in Sec.~\ref{sec:mechanism} as follows.
Let us focus on the dimer limit and look at Fig.~\ref{fig:chains}(b).
Our external magnetic field for the dc case~\eqref{eq:Hext2}
increases the difference between the local magnetic fields
and, hence, the local magnetizations on the left and right sites.
This amounts to the transfer of some positive local magnetization from left to right, resulting in the positive spin current.

The corresponding spin current spectra are shown in Fig,~\ref{fig:mag_master}(c).
For the ac case, there are several harmonic peaks
as discussed in Ref.~\cite{Ikeda2019}.
In particular, the dc component is generated by the spin-current rectification~\cite{Ishizuka2019a}.
As shown in the panel (d), the ac component $|\Jspinmat(\Omega)|$
is proportional to $\beta$, and the dc and second-harmonic components
are to $\beta^2$.
In other words, the results of the ac case
are understood by perturbation in $\beta$.
Whereas the ac output is the linear response,
the dc and second-harmonic ones are the second-order perturbation.

Our finding is that the dc spin current 
for the dc input is proportional to $\beta$
rather than $\beta^2$ as shown in Fig.~\ref{fig:mag_master}.
Thus this dc spin current is significantly larger
for smaller magnetic fields.
Again, the mechanism of the spin current generation is
the one elucidated in Sec.~\ref{sec:mechanism},
and one notices that
the direction of the dc spin current
is reversed by changing the sign of $\beta$.
Therefore, the dc-spin-current direction can be switched
by changing the direction of the external magnetic field.

\section{Discussion and Conclusion}
Studying a simple model of inversion-asymmetric antiferromagnets,
we have proposed the two ways of generating spin currents.
The first one is to utilize an ac electric field,
which leads, through the exchange-interaction modulation, to
the dc and second-harmonic spin currents.
This finding serves as an interesting application of the exchange-interaction control~\cite{Mentink2015}.
The amplitude of the generated spin current by this method
scales as $E_0^2$ with $E_0$ being the amplitude of the input ac electric field.
This second-power scaling in electromagnetic fields are in common with
the different proposals including the spin-current rectification proposed in Refs.~\cite{Ishizuka2019a,Ishizuka2019b}.
Thus the relative importance between these proposals relies on
the prefactors that should depend on the material parameters.

The second way of spin-current generation
is to utilize a dc magnetic field of pulse shape.
In this case, the generated spin current is proportional to
the amplitude of the external magnetic field.
This scaling is better for generating larger spin currents
than the second-power scaling proposed in related studies~\cite{Ishizuka2019a,Ishizuka2019b,Ikeda2019}.
Also, the direction of the spin current can be reversed
by changing the direction of the external magnetic field.
This controllability could be of experimental relevance.

Both ways of spin current generation are understood in a unified manner as in Sec.~\ref{sec:mechanism}.
In inversion-asymmetric antiferromagnets,
there exists some imbalance of local magnetizations at equilibrium.
Once either the exchange interaction or the local magnetic field
is modulated, the magnetization imbalance is converted into spin currents.
We note that this is a transient phenomenon.
In fact, instead of the pulse,
we could turn on the exchange-interaction modulation
and keep it constant for a very long time.
The dc spin current in this situation~\cite{Chinzei} would decrease
as the system approaches a steady state.

The spin current generation mechanism proposed in this paper
is very simple and generic.
This mechanism should contribute to the understanding of spin currents
in antiferromagnets, and its experimental verification is of interest in fundamental and applied physics.
\hl{Upon experimental verifications, one might need
more material-specific models including crystallography and so on}~\cite{Rezende2019}.
\hl{This future direction is of crucial importance in applications.}


\vspace{6pt} 




\section*{Acknowledgments}
Fruitful discussions with K.~Chinzei, M.~Sato, and H.~Tsunetsugu are gratefully acknowledged.
This work was supported by JSPS KAKENHI Grant No. JP18K13495.

\appendix
\section{Methods}\label{sec:mate}

\subsection{fermionization}\label{sec:fermion}

In actual calculations, it is convenient to map
our spin model in Eqs.~\eqref{eq:H0},\eqref{eq:Htot}, and \eqref{eq:Hext}
onto noninteracting spinless fermions.
Following Ref.~\cite{Ikeda2019},
we perform the Jordan-Wigner transformation~\cite{Sachdev2011}:
$\hS^+_j=\prod_{i(<j)}(1-2\cre_i\ann_i)\ann_j$,
$\hS^-_j=\prod_{i(<j)}(1-2\cre_i\ann_i)\cre_j$,
and $\hS^z_j = 1/2- \cre_j \ann_j$,
where $\hS^\pm_j\equiv(\hS^x_j\pm \ii \hS^y_j)/2$
and the creation and annihilation operators satisfy the standard anticommutation relations $\{\ann_i,\cre_j\}=\delta_{ij}$ etc.

Then we simplify our spin model further by
defining the Fourier transforms for the odd and even sites:
$\hat{a}_k \equiv L^{-1/2} \sum_{j=1}^L \ee^{-\ii k (2j)} \ann_{2j}$
and
$\hat{b}_k  \equiv L^{-1/2} \sum_{j=1}^L \ee^{-\ii k (2j+1)} \ann_{2j+1}$,
where $k=\pi m/L$ $(m=0,1,\dots,L-1)$.
The spin Hamiltonians are then mapped to $2\times2$ matrices 
with the following two-component fermion operator:
$\spinor\equiv{}^\text{t}(\hat{a}_k,\hat{b}_k)$.
In fact, one obtains
\begin{align}
	\hzero &= \sum_k \spinor^\dag \hzeromat(k) \spinor;\notag\\
	\hzeromat(k) &=  J\left(\cos k\,\sigma_x -\stagXY \sin k\,\sigma_y\right) -\stagZ \sigma_z,\\
	\hext(t) &= \sum_k \spinor^\dag \hextmat(k,t) \spinor;\notag\\
	\hextmat(k,t) &= J'(t)\left(\cos k\, \sigma_x -\stagXY \sin k\, \sigma_y\right),
\end{align}
where $\sigma_\alpha$ ($\alpha=x,y$, and $z$) are the Pauli matrices.
The spin current~\eqref{eq:sc} is also fermionized as
\begin{align}
	\hJ &= \sum_k \spinor^\dag \Jspinmat(k) \spinor;\notag\\
	\Jspinmat(k) &= -J (\sin k\, \sigma_x + \stagXY \cos k\, \sigma_y).
\end{align}
We remark that the Hamiltonian and the spin current are
represented as sums over $k$'s.
Thus our problem reduces to the direct product of each $k$-subspace.

\begin{figure*}
\includegraphics[width=12cm]{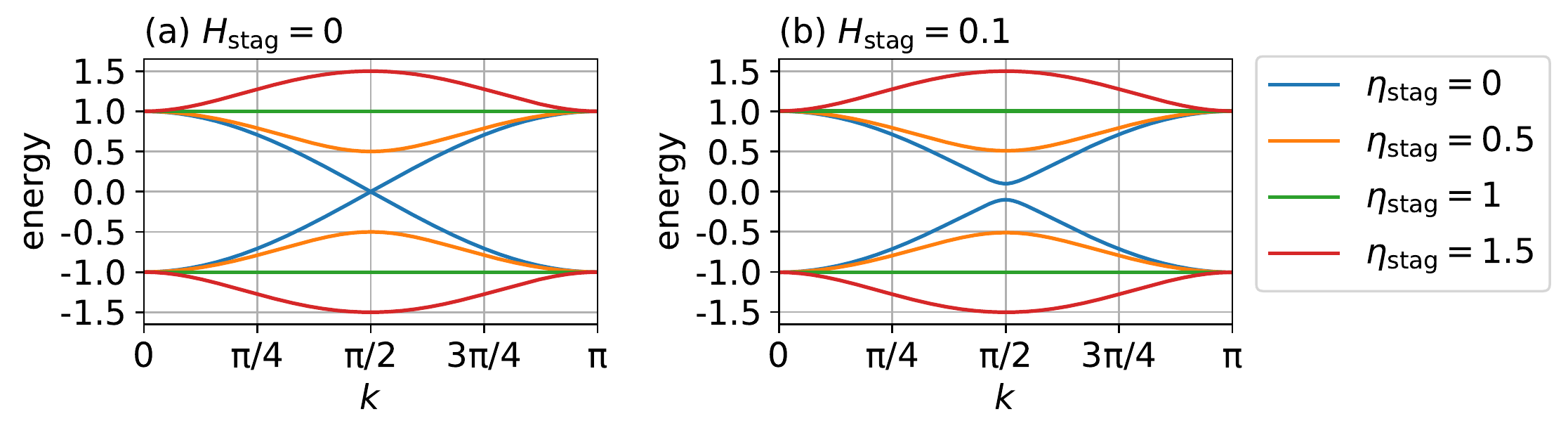}
\caption{Energy bands for four choices of the staggered exchange interaction $\stagXY$
in the absence (a) and presence (b) of the
staggered magnetic field $\stagZ$.
Each parameter is shown in the figure, and $J$ is set to unity.}
\label{fig:bands}
\end{figure*} 

We let $\ket{\phi_\pm(k)}$ denote
the two eigenstates of $\hzeromat(k)$,
\begin{equation}
	\hzeromat(k)\ket{\phi_\pm(k)} = \pm \epI \ket{\phi_\pm(k)}
\end{equation}
with $\epI=\sqrt{J^2 \left[\cos^2 k+(\stagXY\sin k)^2\right]+\stagZ^2}$.
These eigenvalues $\pm\epI$ define the two energy bands,
which are illustrated in Fig.~\ref{fig:bands}.
The band gap is given by $\gapI = 2\sqrt{J^2 \min(1,\stagXY^2) +\stagZ^2}$.
The ground state of the total system is such a state that
all the states in the lower (upper) band are occupied (unoccupied).
Note that this ground state is half-filled
and, hence, has zero magnetization in the spin language.

\subsection{quantum master equation}
We have analyzed the dynamics under pulse fields
by using the quantum master equation (see Ref.~\cite{Ikeda2019} for more detail):
\begin{align}
	\frac{\dd}{\dd t}\rho(k,t)
	&=-\ii [H(k,t),\rho(k,t)] + \mathcal{D}[\rho(k,t)],\label{eq:QME}\\
	\mathcal{D}[\rho(k,t)]&\equiv\gamma\left(L_k\rho(k,t) L_k^\dag -\frac{1}{2}\{ L_k^\dag L_k,\rho(k,t)\}\right),
\end{align}
where $\rho(k,t)$ is the $2\times2$ reduced density matrix for the $k$-subspace.
In the Hamiltonian matrix $H(k,t)=\hzeromat(k)+\hext(k,t)$,
the time-dependent exchange interaction
is replaced by the pulse one as Eq.~\eqref{eq:Jpulse}.
The first term on the right-hand-side of Eq.~\eqref{eq:QME}
is the same as the time-dependent Schr\"{o}dinger equation
whereas the second one describes the relaxation effect.
The Lindblad operator $L_k\equiv\ket{\phi_-(k)}\bra{\phi_+(k)}$
causes the excited state $\ket{\phi_+(k)}$ relaxing to the ground state $\ket{\phi_-(k)}$
at rate $\gamma$~\cite{Breuer2007}.
\hl{Thus our model corresponds to the case that the system is in contact with a reservoir at zero temperature,
and thermal fluctuations are neglected.
}

Our master equation~\eqref{eq:QME} is a set of ordinary differential equations.
Thus we can numerically solve it by an explicit method
such as the Runge-Kutta method.

Finally, we remark that the continuity equation
$d \hat{S}^z_i/dt + \hlocj_i - \hlocj_{i-1}=0$
is corrected by the $\mathcal{D}$ term in the master equation.
To see this, we consider the full master equation $\dd \hat{\rho}/\dd t = -\ii [\htot(t),\hat{\rho}]+\hat{\mathcal{D}}(\hat{\rho})$ instead of Eq.~\eqref{eq:QME} reduced to each $k$.
Then the expectation value $S_j^z(t)=\text{tr}(\hat{\rho}(t)\hS_j^z)$ satisfies the following equation
\begin{align}
	\frac{\dd S_j^z(t)}{\dd t} + j_i^{\text{spin}}(t)-j_{i-1}^{\text{spin}}(t) = \text{tr}\{\hat{\mathcal{D}}[\hat{\rho}(t)] \hS_j^z\}.
\end{align}
The right-hand side gives the source or sink for the magnetization
and, hence, the standard continuity equation does not hold true.
Note that this is not a contradiction because the magnetization,
or the angular momentum along a certain direction,
is not conserved in general
unlike the electric charge that is strictly conserved.



%

\end{document}